\documentclass[conference]{IEEEtran}
\IEEEoverridecommandlockouts

\ifCLASSINFOpdf
  \usepackage[pdftex]{graphicx}
\else
  \usepackage[dvips]{graphicx}
\fi

\usepackage{amsmath}
\usepackage{tabularx}
\usepackage{algorithm}
\usepackage{algorithmic}
\usepackage{subfig} 
\usepackage{float} 
\usepackage{url}
\usepackage{flushend}

\newlength{\figureWidth} 
\begin{document}

\title{Evolutionary Algorithm for Graph Anonymization}

\author{
 \IEEEauthorblockN{Jordi Casas-Roma}
 \IEEEauthorblockA{
 Universitat Oberta de Catalunya\\
 Barcelona, Spain\\
 jcasasr@uoc.edu }
 \and
 \IEEEauthorblockN{Jordi Herrera-Joancomart\'{\i} }
 \IEEEauthorblockA{
 Universitat Aut\`{o}noma de Barcelona\\
 Bellaterra, Spain\\
 jherrera@deic.uab.cat }
 \and
 \IEEEauthorblockN{Vicen\c{c} Torra }
 \IEEEauthorblockA{Artificial Intelligence Research Institute\\
 Spanish National Research Council\\
 Bellaterra, Spain\\
 vtorra@iiia.csic.es}

\thanks{This paper is an updated and revised version of the paper published in Casas-Roma, J., Herrera-Joancomart\'{i}, J. and Torra, V. (2012). Algoritmos gen\'{e}ticos para la anonimizaci\'{o}n de grafos. In XII Reuni\'{o}n Espa\~{n}ola sobre Criptolog\'{i}a y Seguridad de la Informaci\'{o}n (RECSI 2012) (pp. 243-248). Donostia-San Sebasti\'{a}n.}
 
\thanks{\textbf{Acknowledgments:} This work was partially supported by the Spanish MCYT and the FEDER funds under grants TSI2007-65406-C03 ``E-AEGIS'', TIN2010-15764 ``N-KHRONOUS'', CONSOLIDER CSD2007-00004 ``ARES'', and TIN2011-27076-C03 ``CO-PRIVACY''.}
}

\maketitle

\begin{abstract}
In recent years, there has been a significant increase in the use of graph-formatted data. Socials networks, among others, represent relationships among users and present interesting information for researches and other third-parties. The problem appears when someone wants to publicly release this information, especially in the case of social or healthcare networks. In these cases, it is essential to implement an anonymization process in the data in order to preserve the privacy of users who appears in the network. In this paper we present an algorithm for graph anonymization, called Evolutionary Algorithm for Graph Anonymization (EAGA), based on edge modifications to preserve the $k$-anonymity model.
\end{abstract}

\section{Introduction}
\label{sec:intro}

In recent years, the representation of data on graph format has experienced an exponential growth. This data format allows the representation of complex structures in an easier way than the traditional relational data. In graph-formatted data each node represents a user or an entity, with some optional number of numerical, nominal or categorical attributes. In addition, the edges or links between nodes stand for relationships among users or entities in a richer and more intuitive way. A good example is presented by social networks. Regardless of their kind or target, social networks have a lot of interesting information for different field studies (psychology, social sciences, etc). Therefore, graph data is very interesting to scientists and companies around the world. The problem emerges with the need to preserve the privacy of individuals who appear in these social networks.

\begin{figure}[h]
	\centering
	\subfloat[$G$]{\label{fig:naive-anonimization-1a}\includegraphics[width=3.65cm]{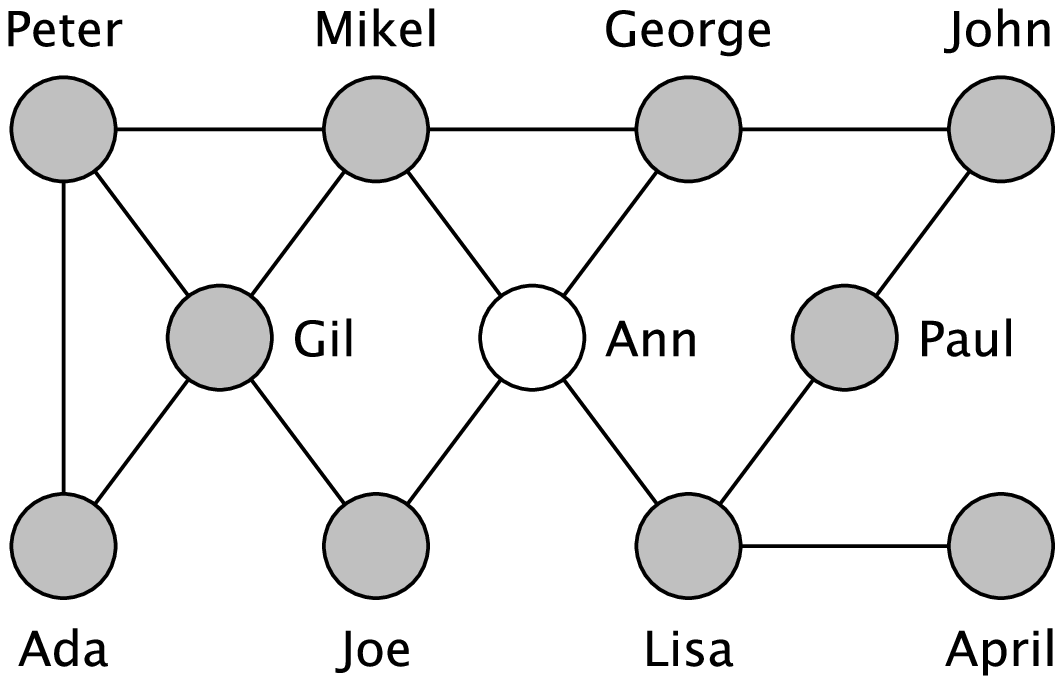}}
	\subfloat[$G_{1}$]{\label{fig:naive-anonimization-1b}\includegraphics[width=3.65cm]{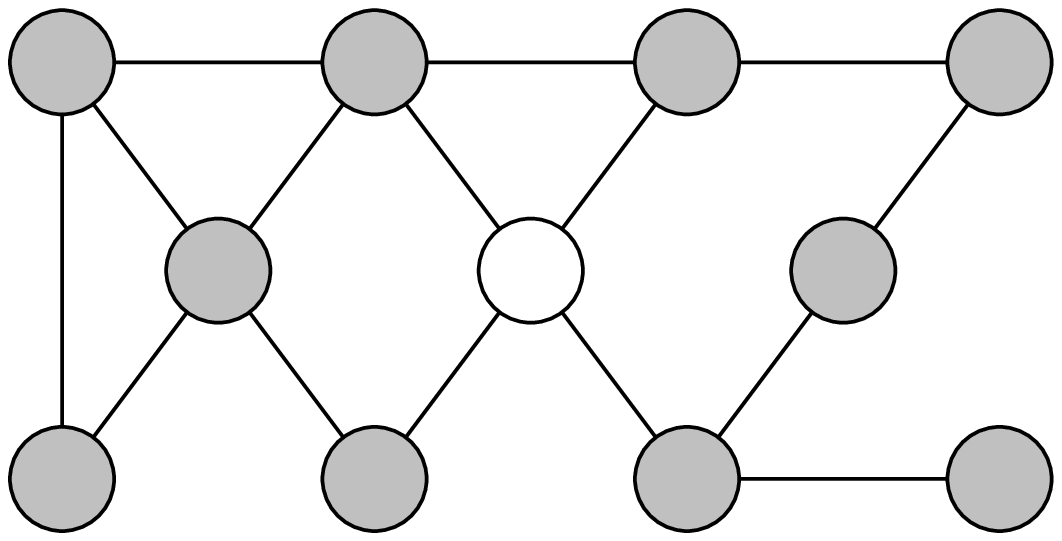}}
	\subfloat[$G_{Ann}$]{\label{fig:naive-anonimization-1c}\includegraphics[width=1.69cm]{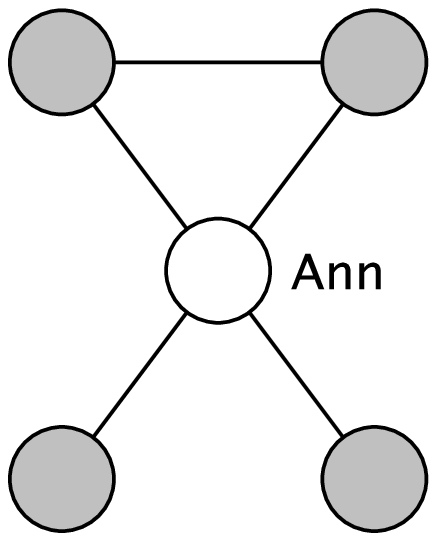}}
	\caption{Anonymization example, where $G$ is the original graph, $G_{1}$ is the na\"{i}ve anonymous version and $G_{Ann}$ is 1-neighbourhood of Ann.}
	\label{fig:naive-anonimization-1}
\end{figure}

A preliminary approach to data anonymization, known as na\"{i}ve anonymization, consists of removing all attribute-based information which allows an attacker to re-identify the user in the anonymized graph. An example is shown in Figure \ref{fig:naive-anonimization-1}.

Figure \ref{fig:naive-anonimization-1a} shows a toy example of a social network, where each node represents an individual and each edge indicates the friendship relation between them. Figure \ref{fig:naive-anonimization-1b} presents the same graph after a na\"{i}ve anonymization, where node identifiers have been removed and the graph structure remains the same. One can think users' privacy is safe, but an attacker can break the privacy and re-identify a user on an anonymized graph. For instance, if an attacker knows that Ann has four friends and two of them are friends themselves, then the adversary can construct the 1-neighbourhood of Ann, depicted in Figure \ref{fig:naive-anonimization-1c}. From this sub-graph, the attacker can uniquely re-identify user Ann on anonymized graph. Consequently, user's privacy has been broken by the attacker.

The example we have depicted in Figure \ref{fig:naive-anonimization-1} is quite simple, but it gives us an idea of the complexity on graph anonymization process. Apart from the user's properties, the structure and relation among users can be used by an attacker to re-identify users and attack the privacy of the graph.

In this paper we present an algorithm, called \textit{Evolutionary Algorithm for Graph Anonymization} (EAGA), in order to preserve the user's privacy on graphs, and it is based on the $k$-anonymity model.

This paper is organized as follows. The notation is summarized in Section \ref{sec:notation}. In Section \ref{sec:stateOfTheArt}, we review the state of the art. Our algorithm is presented in Section \ref{sec:genetic-algorithm}.  In Section \ref{sec:setup} we point out our experimental set up and then, we discuss empirical results in Section \ref{sec:results}. Lastly, we outline the conclusions and propose guidelines for future work in Section \ref{sec:conclusions}.

\section{Notation}
\label{sec:notation}

Let $G=(V,E)$ be a simple graph, where $V$ is the set of nodes and $E$ the set of edges in $G$. We define $n=\vert V \vert$ to denote the number of nodes and $m=\vert E \vert$ to denote the number of edges. We use $d$ to define the degree sequence of $G$, where $d=\{d_1, d_2, \dots, d_n\}$ is a vector of length $n$ and $d_i$ is the degree of node $v_{i} \in V$. We use $\Gamma(v_i)$ to denote the neighbourhood of node $v_i$, i.e, nodes adjacent to node $v_i$. 

\section{State of the Art}
\label{sec:stateOfTheArt}

There are different approaches for graph anonymization, such as randomization methods (for example \cite{HayEtAl:2007}, \cite{YingEtAl:2009} and \cite{ZhangZhang:2009}), which are based on random edge modification processes, or generalization methods (for instance \cite{HayEtAl:2008} and \cite{CampanTruta:2009}), which are based on node and edge clustering to construct super-nodes and super-edges hiding the individuals' properties. Nevertheless, we will focus on $k$-anonymity methods in this paper. The concept of $k$-anonymity was introduced by Sweeney \cite{Sweeney:2002} for the privacy preservation on relational data. Formally, the $k$-anonymity model is defined as: let $RT(A_{1},\ldots,A_{n})$ be a table and $QI_{RT}$ be the quasi-identifier associated with it. $RT$ is said to satisfy $k$-anonymity if and only if each sequence of values in $RT\left[QI_{RT}\right]$ appears with at least $k$ occurrences in $RT\left[QI_{RT}\right]$. The $k$-anonymity model indicates that an attacker can not distinguish between different $k$ records although he manages to find a group of quasi-identifiers. Therefore, the attacker can not re-identify an individual with a probability greater than $\frac{1}{k}$.

Different concepts can be used to apply the $k$-anonymity model on graphs. A widely option is using the node degree as a quasi-identifier \cite{LiuTerzi:2008}. It is called $k$-degree anonymity. We assume that the attacker knows the degree of some target nodes. If the attacker identifies a single node with equal degree in the anonymized graph, then he has re-identified this node. $K$-degree anonymous methods are based on modifying the graph structure (by adding and removing edges) to ensure that all nodes satisfy the $k$-anonymity. In other words, the main objective is that all nodes have at least $k-1$ other nodes sharing the same degree.

Pei and Zhou \cite{ZhouPei:2008} consider as quasi-identifier the 1-neighbourhood sub-graph of the objective nodes. Let $k$ be a positive integer. For a vertex $v_i \in V$, $v_i$ is $k$-anonymous in $G$ if there are at least $k-1$ other vertices $v_1, \ldots, v_{k-1} \in V$ such that $\Gamma(v_i), \Gamma(v_1), \ldots, \Gamma(v_{k-1})$ are isomorphic. $G$ is $k$-anonymous if every vertex is $k$-anonymous in $G$. It is called $k$-neighbourhood anonymity. Zhou et al. \cite{ZouEtAl:2009} consider all structural information about a target node as quasi-identifier and propose a new model called $k$-automorphism to anonymize a graph and ensure privacy against this attack. They define a $k$-automorphic graph as follows: given a graph $G$, if $k-1$ automorphic functions $F_{a} (a=1, \ldots, k-1)$ exist in $G$, and for each vertex $v$ in $G$, $F_{a_1}(v) \neq F_{a_2} (1 \leq a_1 \neq a_2 \leq k-1)$, then $G$ is called a $k$-automorphic graph. Hay et al. \cite{HayEtAl:2008} go a step further. They propose a method, named $k$-candidate anonymity, which uses queries as quasi-identifier. In this method, a node $v_{i}$ is $k$-candidate anonymous with regard to question $Q$ if there are at least $k-1$ other nodes in the graph with the same answer. Formally, $\vert cand_{Q}(v_{i}) \vert \geq k$ where $cand_{Q}(v_{i})=\{ v_{j} \in V \vert Q(v_{j})=Q(v_{i}) \} $. A graph is $k$-candidate anonymous to question $Q$ if all of its nodes are $k$-candidate anonymous concerning question $Q$. The question $Q$ is modelled according to assume adversary's knowledge.

\subsection{$k$-degree anonymity}

Liu and Terzi \cite{LiuTerzi:2008} develop a method based on adding and removing edges from the original graph $G=(V,E)$ in order to construct a new graph $\widetilde{G}=(\widetilde{V}, \widetilde{E})$ which fulfils $k$-degree anonymity model. In $k$-degree anonymity we presume that the only possible attack is when the attacker knows the degree of some vertices. Therefore, if some vertex is identified with certainty with this information, then we have an information leakage. Formally, an anonymous graph $\widetilde{G}=(\widetilde{V},\widetilde{E})$ has to verify the following restrictions: (1) $\widetilde{G}$ must be $k$-degree anonymous, $V=\widetilde{V}$ and $E \cap \widetilde{E} \approx E$. This model ensures the graph against re-identification attacks based on degree knowledge of the adversary. Usually, the bigger the $k$ value, the bigger the privacy and also the information loss.

Their algorithm is two-step based. The first one modifies the degree sequence of the original graph. The authors seek how to obtain an anonymous $k$-degree sequence for a given specific $k$ value with the minimum number of changes on the degree sequence. They resolve this part through linear programming techniques. Then, the second step constructs a new graph $\widetilde{G}_{0}$ from the anonymized $k$-degree sequence generated on first step. Then, they apply edge swap iteratively in order to obtain a graph as equal as possible to the original one. The edge swap is an operation among four nodes $v_{a}, v_{b}, v_{c}, v_{d} \in \widetilde{G}_{i}=(V,\widetilde{E}_{i})$ where $(v_a, v_c), (v_b, v_d) \in \widetilde{E}_{i}$ and $(v_a, v_b), (v_c, v_d) \notin \widetilde{E}_{i}$ or $(v_a, v_d), (v_b, v_c) \notin \widetilde{E}_{i}$ where $\widetilde{G}_{i}=(V, \widetilde{E}_{i})$ is the graph $\widetilde{G}_{0}$ after $i$ iterations. The target is to achieve an edge set as similar as possible to the original one ($E \cap \widetilde{E} \approx E$). 

\section{\textit{EAGA Algorithm}}
\label{sec:genetic-algorithm}

In this section we will present our approach for graph anonymization, called Evolutionary Algorithm for Graph Anonymization (EAGA)\footnote{Source code available at: \url{http://deic.uab.cat/~jcasas/}}, which is based on evolutionary algorithms and focused on creating a $k$-degree anonymous graph.

A high-level description of our proposal allows us to structure our anonymization algorithm in two steps, similar to the approach by Liu and Terzi \cite{LiuTerzi:2008}:

\begin{enumerate}
	\item In the first step, from the original degree sequence of $G=(V,E)$, $d=\{d_1, \cdots ,d_n\}$, we construct a new sequence $\widetilde{d}$ which is $k$-degree anonymous and minimize the distance $\Delta$ from the original sequence computed by Equation \ref{eq:distance-1}.
	
\begin{equation}
\label{eq:distance-1}
\Delta(\widetilde{d}, d) = \sum_{i=0}^{n} \vert \widetilde{d_{i}} - d_{i} \vert
\end{equation}	

	\item In the second step, we construct a graph $\widetilde{G}=(\widetilde{V},\widetilde{E})$ where $\widetilde{V}=V$, $\widetilde{E} \cap E \approx E$ and the degree sequence is equal to $\widetilde{d}$.
\end{enumerate}

The process of creating the $k$-anonymous degree sequence determines the anonymization level and the distance from the original degree sequence. An optimal sequence has to provide the requested $k$-anonymity level and has to minimize the distance from the original degree sequence. This last condition is decisive for data utility and information loss. 

\subsection{Step I: Obtaining the $k$-degree anonymous sequence}

The problem of obtaining a $k$-anonymous degree sequence has certain peculiarities that must be considered:
\begin{itemize}
	\item The number of elements in the degree sequence determines the number of nodes. Therefore, this value cannot be altered.
	\item The values of the degree sequence are the degree of the nodes. Hence, these values have to be integer in range $[0, n-1]$.
	\item The total number of edges is half the sum of the degree sequence, since each edge is counted twice in the degree sequence. To preserve the number of edges, the sum of the anonymized sequence must be equal to the sum of the original sequence, i.e, $\sum_{i=0}^{n} \widetilde{d}_i = \sum_{i=0}^{n} d_i$.
	\item Each change on the degree sequence has to be translated as an edge modification into the anonymous graph. Thus, it is necessary to perform the minimum number of changes in the degree sequence (minimizing the distance between the two sequences) to obtain an anonymous graph with the minimum number of changes from the original one.
\end{itemize}

Our proposal uses evolutionary algorithms to generate the $k$-degree anonymous sequence. Algorithm \ref{alg:EAGA-1} details the above steps to generate an anonymous degree sequence.

\begin{algorithm}
	\caption{Algorithm pseudo-code for generating $k$-degree anonymous sequence.}
	\label{alg:EAGA-1}
	\begin{algorithmic}
		\REQUIRE{Original degree sequence ($d$) and the $k$-anonymity value ($k$).}
		\ENSURE{$k$-degree anonymous sequence ($\widetilde{d}$).}

\STATE INITIALIZE $population \Leftarrow d$
\STATE $k\_actual \Leftarrow$ GET\_K $population$
\WHILE {$k\_actual < k$}
	\STATE MUTATE $population$
	\STATE EVALUATE $new$ $candidates$
	\STATE $population \Leftarrow$ SELECT $individuals$
	\STATE $k\_actual \Leftarrow$ GET\_K $individuals$
\ENDWHILE
\STATE $\widetilde{d} \Leftarrow$ SELECT $best$ $candidate$
\RETURN $\widetilde{d}$

	\end{algorithmic}
\end{algorithm}

As we have shown in Algorithm \ref{alg:EAGA-1}, population is initialized from original degree sequence. Next, the sentences in the \textit{while} loop are the generation step. Here we apply the basic mutation process (MUTATE function in Algorithm \ref{alg:EAGA-1}) which adds one to an element of the sequence and subtracts one to another element of the sequence. This operation represents edge swap, which is the most basic edge modification on a graph. For example, if an edge $(v_0, v_1)$ is modified by replacing one node, one can obtain $(v_0, v_2)$. This edge modification is represented on the degree sequence as a subtraction on node $v_1$ (because it decreases its degree) and a addition on node $v_2$ (because it increases its degree). It is important to note that our algorithm does not use crossover since this operation systematically breaches the rule that preserves the number of edges of the graph, generating invalid candidates. We consider the performance of the algorithm would be affected by the inclusion of this type of evolution, and improvements would not occur in time or quality of the solution found.

When candidate generation is done, we evaluate the candidates in order to find the best one. The score of each candidate is determined by the fitness function (EVALUATE function in Algorithm \ref{alg:EAGA-1}). This function assigns different score punctuation whether each individual fulfils the desired $k$ or not. Individuals who do not meet the desired $k$-anonymity value are scored in range [0,1] considering two parameters:

\begin{itemize}
	\item The number of nodes which do not fulfil the $k$-anonymity.
	\item The dispersion level, computed as the average distance from all nodes to the mean degree value.
\end{itemize}

Contrary, individuals who fulfil the desired privacy level are scored in range [1,2] considering only one parameter:
\begin{itemize}
	\item The distance from the original sequence. The target is to minimize this value, as we described in Equation \ref{eq:distance-1}
\end{itemize}

Finally, the candidate selection uses the \textit{steady-state} model. According to it, the worst candidates of the actual generation are replaced by the best candidates of the new generation.

\subsection{Step II: Modifying the original graph}

The result of the first step is a $k$-degree anonymous sequence. Then, in the second step we apply the necessary modifications to original graph in order to obtain the $k$-anonymous one. The anonymized $k$-degree sequence informs the degree for each node on anonymized graph. Therefore, the difference between original and anonymized degree sequence points to nodes which have to increase or decrease their degree. Hence, we have to add or remove edges to/from these nodes.

As we can see in Algorithm \ref{alg:EAGA-2}, this step begins computing the difference vector, $d_{dif} = d - \widetilde{d}$, which allows us to easily detect which nodes have to increase or decrease their degree. The algorithm removes incident edges to nodes which have to decrease their degree, while it adds new edges to nodes which have to increase their degree. We apply these modifications removing the edge $(v_p, v_q) \in E$, where $v_{q}$ belongs to nodes which have to decrease their degree, and adding a new edge $(v_p, v_r)$, where $v_{r}$ belongs to nodes which have to increase their degree.

\begin{algorithm}
	\caption{Algorithm pseudo-code for modifying the original graph.}
	\label{alg:EAGA-2}
	\begin{algorithmic}
		\REQUIRE{Original graph $G(V,E)$, original degree sequence $d$ and the $k$-degree anonymous sequence $\widetilde{d}$.}
		\ENSURE{The graph $\widetilde{G}(V,\widetilde{E})$ where the degree sequence is $\widetilde{d}$ and $\widetilde{E} \cap E \approx E$.}

\STATE $\widetilde{G}(V,\widetilde{E}) \Leftarrow \widetilde{G}_{0}(V,\widetilde{E})$
\STATE $d_{dif} = d - \widetilde{d}$
\STATE $V_{del} = \{ v_{i} \in V \vert d_{dif}(i) < 0 \}$
\STATE $V_{add} = \{ v_{i} \in V \vert d_{dif}(i) > 0 \}$
\WHILE{$V_{del} \neq \emptyset$ and $V_{add} \neq \emptyset$}
\STATE $\widetilde{E} = \widetilde{E} \setminus \{ (v_p, v_q) \}$ where $(v_p, v_q) \in E$ and $v_{q} \in V_{del}$
\STATE $V_{del} = V_{del} \setminus \{ v_{q} \}$
\STATE $\widetilde{E} = \widetilde{E} \cup \{ (v_p, v_r) \}$ where $v_{r} \in V_{add}$
\STATE $V_{add} = V_{add} \setminus \{ v_{r} \}$
\ENDWHILE
\RETURN $\widetilde{G}$

	\end{algorithmic}
\end{algorithm}

\section{Experimental Set Up}
\label{sec:setup}

Three real networks have been used to test the \textit{EAGA} algorithm: \textit{Zachary's Karate Club} \cite{Zachary:1977}, \textit{American College Football} \cite{GirvanNewman:2002} and \textit{Jazz Musicians} \cite{GleiserDanon:2003}. Table \ref{table:datasets-properties-1} presents a summary of their properties.

\begin{table}[h]
	\centering{}
	\begin{tabular}{ | l || r | r | r | r | r | }
		\hline
		\textit{Datasets} & \textit{Nodes} & \textit{Edges} & \textit{Av.deg.} & \textit{Av.dist.} & \textit{Diam.} \\
		\hline
		\hline
		Zachary's Karate Club 	& 34 	& 78 		& 4.588 	& 2.408 	& 5 \\
		\hline
		American College 		& 115 	& 613 		& 10.661 	& 2.508 	& 4 \\
		\hline
		 Jazz Musicians 			& 198 	& 2,742 	& 27.697 	& 2.235 	& 6 \\
		\hline
	\end{tabular}
	\caption{Summary of selected network properties: Number of nodes (\textit{Nodes}), Number of edges (\textit{Edges}), Average degree (\textit{Av. deg.}), Average distance (\textit{Av. dist.}) and diameter (\textit{Diam.})}
	\label{table:datasets-properties-1}
\end{table}

For each dataset, we analyse the evolution of the degree histogram, comparing the histogram on anonymized and original graphs. We use edge intersection to quantify the number of edges which were on original graph and still are on anonymized graph. Clearly, the higher the value, the less the perturbation. This measure is defined by Equation \ref{eq:edgeintersection} as follows:

\begin{equation}
EI(G,\widetilde{G}) = \frac{\vert E \cap \widetilde{E} \vert}{max(\vert E \vert, \vert \widetilde{E} \vert)}
\label{eq:edgeintersection}
\end{equation}

We also analyse three measures related to node centrality in order to quantify the perturbation introduced on anonymized data. The first measure is betweenness centrality. It measures the fraction of number of shortest paths that go through each vertex. Formally, we define the betweenness centrality of node $v_{i}$ as:

\begin{equation}
BC(v_{i}) = \frac{1}{n^{2}} \sum_{s,t} \frac{g^{i}_{st}}{g_{st}}
\end{equation}

\noindent where $g^{i}_{st}$ is the number of geodesic paths from $v_s$ to $v_t$ that pass through $v_i$, and $g_{st}$ is the total number of geodesic paths from $v_s$ to $v_t$.

Closeness centrality is the second centrality measure we have used. It is defined as the inverse of the average distance to all accessible nodes. Formally, we define the closeness centrality of a node $v_{i}$ in Equation \ref{eq:closenesscentrality}.

\begin{equation}
CC(v_{i}) = \frac{n}{\sum_{j} d_{ij}}
\label{eq:closenesscentrality}
\end{equation}

Finally, the third one is degree centrality, which evaluates the centrality of each node associated with its degree. We define the degree centrality of a node $v_{i}$ in Equation \ref{eq:degreecentrality} as follows:

\begin{equation}
DC(v_{i}) = \frac{\Gamma(v_{i})}{m}
\label{eq:degreecentrality}
\end{equation}

The centrality measures described above evaluate the centrality of each node of the graph from different concepts of centrality. These measures give us a value of centrality for each node. To assess the perturbation introduced in the whole graph, we compute the vector of differences for each node between the original and the anonymous graph and then, compute the root mean square (RMS) to obtain a single value for the entire graph. We define the difference of the centrality measure between the original and the anonymous graph as:

\begin{equation}
Dif(G,\widetilde{G}) = \sqrt{\frac{1}{n} ((g_{1} - \widetilde{g}_{1})^{2} + \ldots + (g_{n} - \widetilde{g}_{n})^{2})}
\label{eq:diff}
\end{equation}

\noindent where $g_{i}$ is the value of the centrality measure for node $v_i$ of $G$ and $\widetilde{g}_{i}$ is the value of the centrality measure for node $v_i$ of $\widetilde{G}$. In our experiments we use Equation \ref{eq:diff} to compute a value representing the error induced in the whole graph by EAGA.

Lastly, we want to evaluate how node set evolves during the process of anonymization. In order to do this, we use the Vertex Refinement Queries \cite{HayEtAl:2007} \cite{HayEtAl:2008}. This type of queries models the local neighbourhood structure of a node in the graph. The weakest knowledge query, $\mathcal{H}_{0}(v_{j})$, simply returns the label of the node $v_{j}$. The queries are successively more descriptive: $\mathcal{H}_{1}(v_{j})$ returns the degree of $v_{j}$, $\mathcal{H}_{2}(v_{j})$ returns the list of each neighbours' degree, and so on. The queries can be defined iteratively, where $\mathcal{H}_{i}(v_{j})$ returns the multi-set of values which are the result of evaluating $\mathcal{H}_{i-1}$ on the set of nodes adjacent to $v_{j}$:

\begin{equation}
\mathcal{H}_{i}(v_{j}) = \{ \mathcal{H}_{i-1}(v_{1}), \mathcal{H}_{i-1}(v_{2}), \ldots , \mathcal{H}_{i-1}(v_{p}) \}
\end{equation}

\noindent where $v_{1}, v_{2}, \ldots , v_{p}$ are the nodes adjacent to $v_{j}$.

A candidate set for a query $\mathcal{H}_{i}$ ($cand_{\mathcal{H}_{i}}$) is the set of all nodes with the same value of $\mathcal{H}_{i}$. Consequently, the cardinality of the candidate set for $\mathcal{H}_{i}$ is the number of indistinguishable nodes in $G$ under $\mathcal{H}_{i}$. Note that if the cardinality of the smallest candidate set under $\mathcal{H}_{1}$ is $k$, the probability of re-identification is $\frac{1}{k}$. Hence, the $k$-degree anonymity value for $G$ is $k$.

\begin{equation}
cand_{\mathcal{H}_{1}} = \{v_{j} \in V \vert \mathcal{H}_{1}(v_{i}) = \mathcal{H}_{1}(v_{j}) \}
\end{equation}

We use the $cand_{\mathcal{H}_1}$ to analyse the evolution of nodes, in terms of $k$-degree anonymity, during the anonymization process.

\section{Experimental results}
\label{sec:results}

\begin{figure*}[!ht]
	\centering
	\subfloat[Degree histogram (Zachary's Karate Club)]{\label{fig:karate-1}\includegraphics[width=\figureWidth]{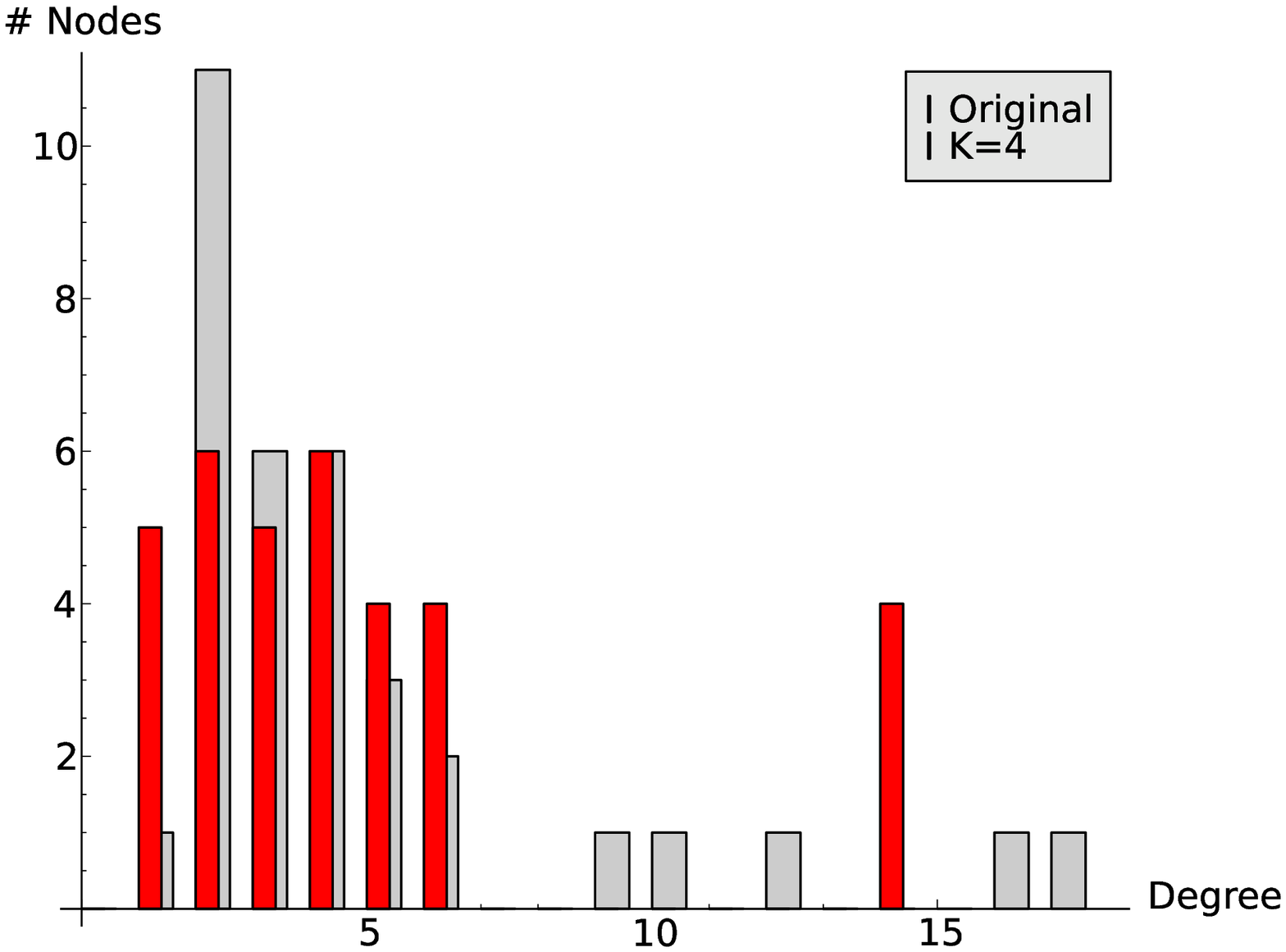}}~
	\subfloat[American College]{\label{fig:football-1}\includegraphics[width=\figureWidth]{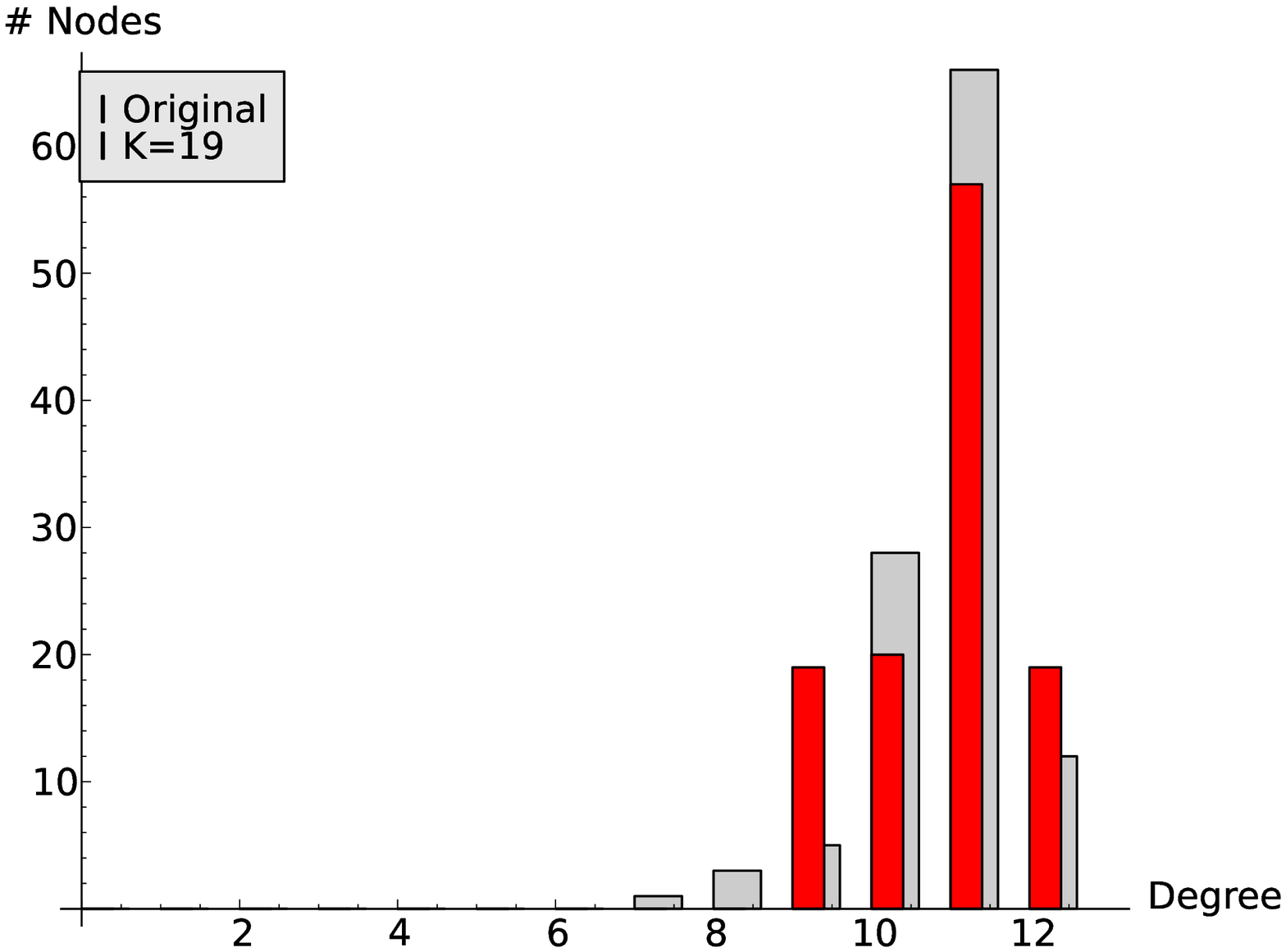}}~
	\subfloat[Jazz Musicians]{\label{fig:jazz-1}\includegraphics[width=\figureWidth]{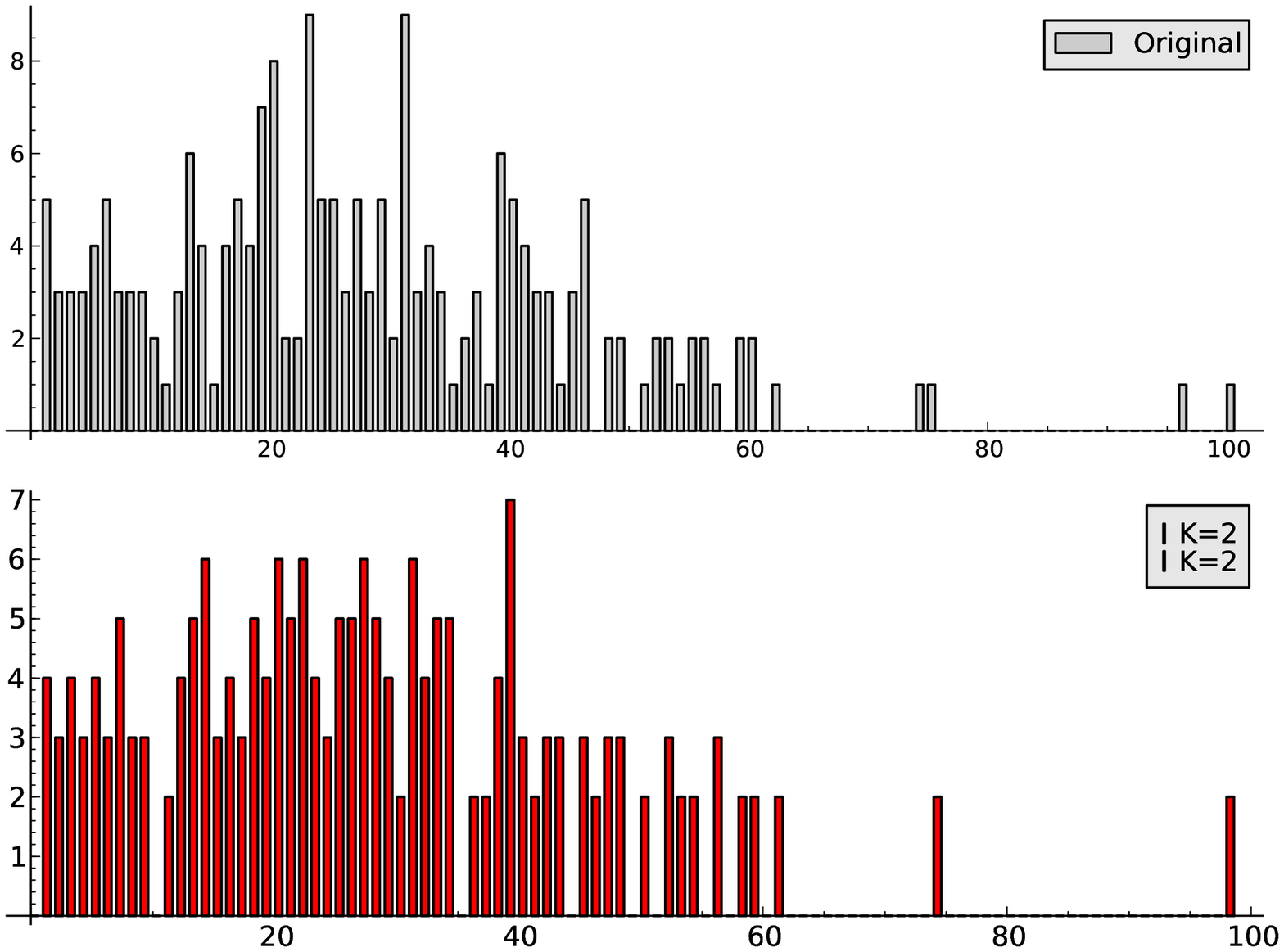}}~	
	\\
	\subfloat[Edge intersection (Zachary's Karate Club)]{\label{fig:karate-3}\includegraphics[width=\figureWidth]{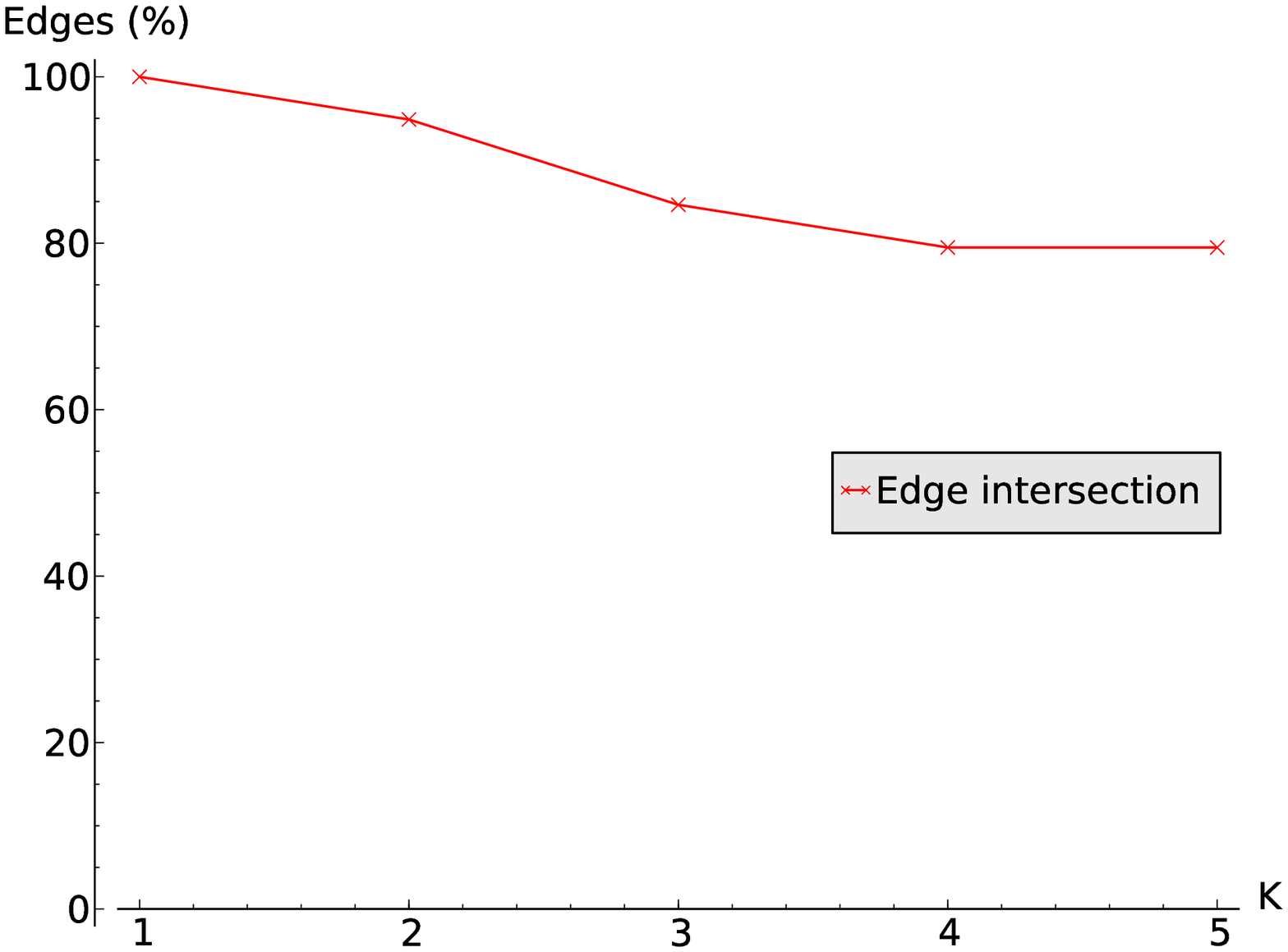}}~
	\subfloat[American College]{\label{fig:football-3}\includegraphics[width=\figureWidth]{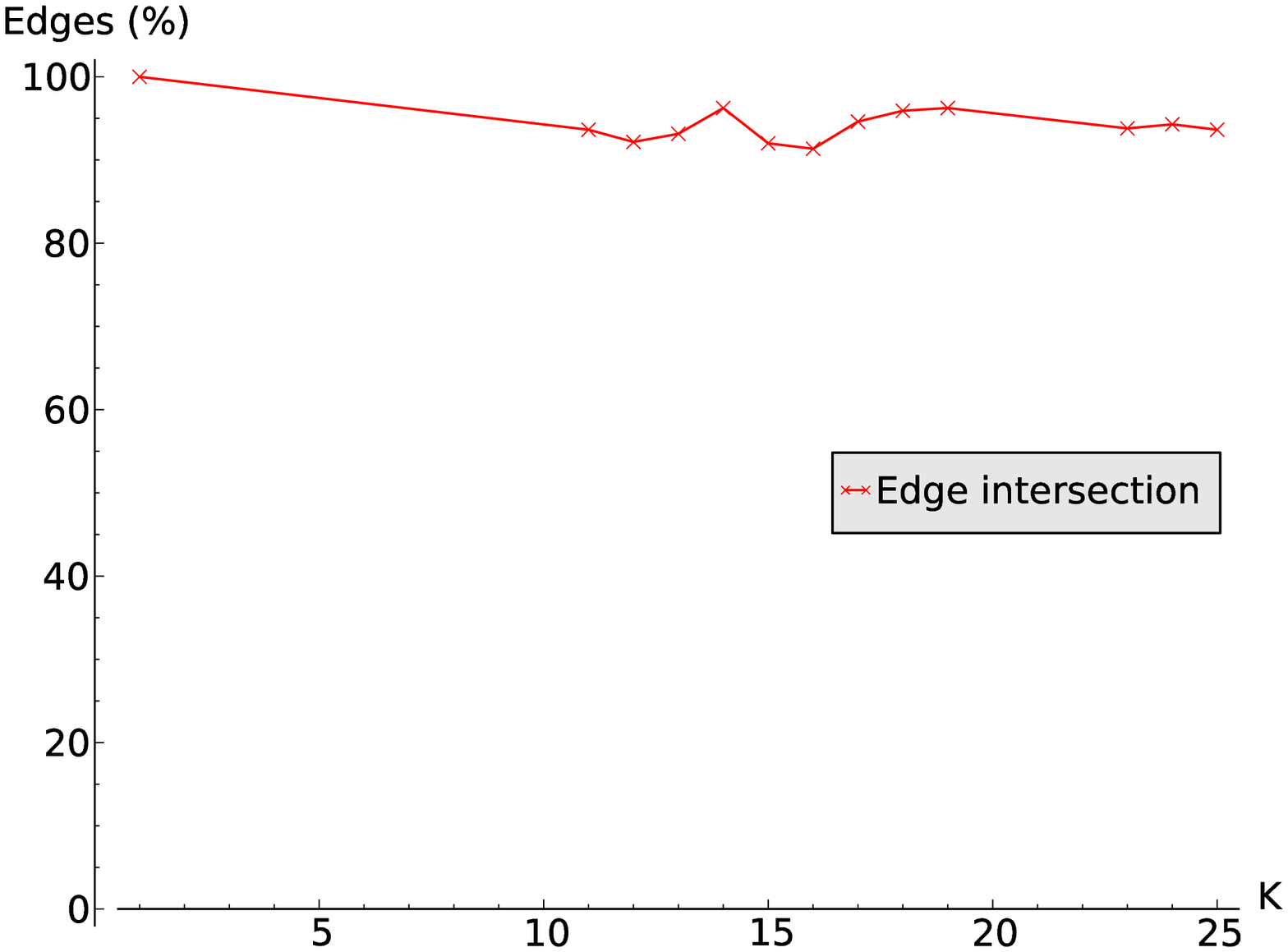}}~
	\subfloat[Jazz Musicians]{\label{fig:jazz-3}\includegraphics[width=\figureWidth]{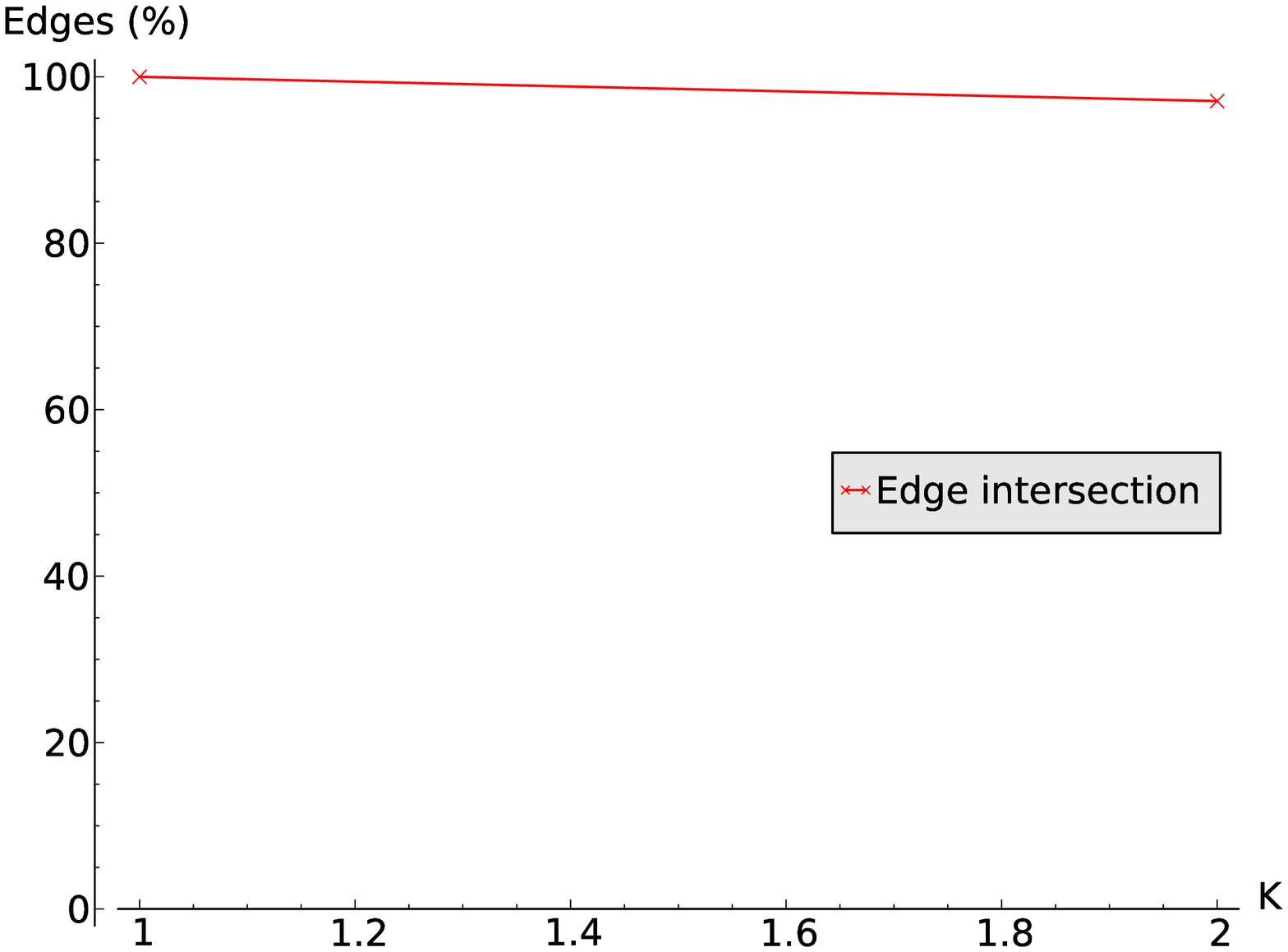}}~	
	\\		
	\subfloat[Centrality RMS (Zachary's Karate Club)]{\label{fig:karate-2}\includegraphics[width=\figureWidth]{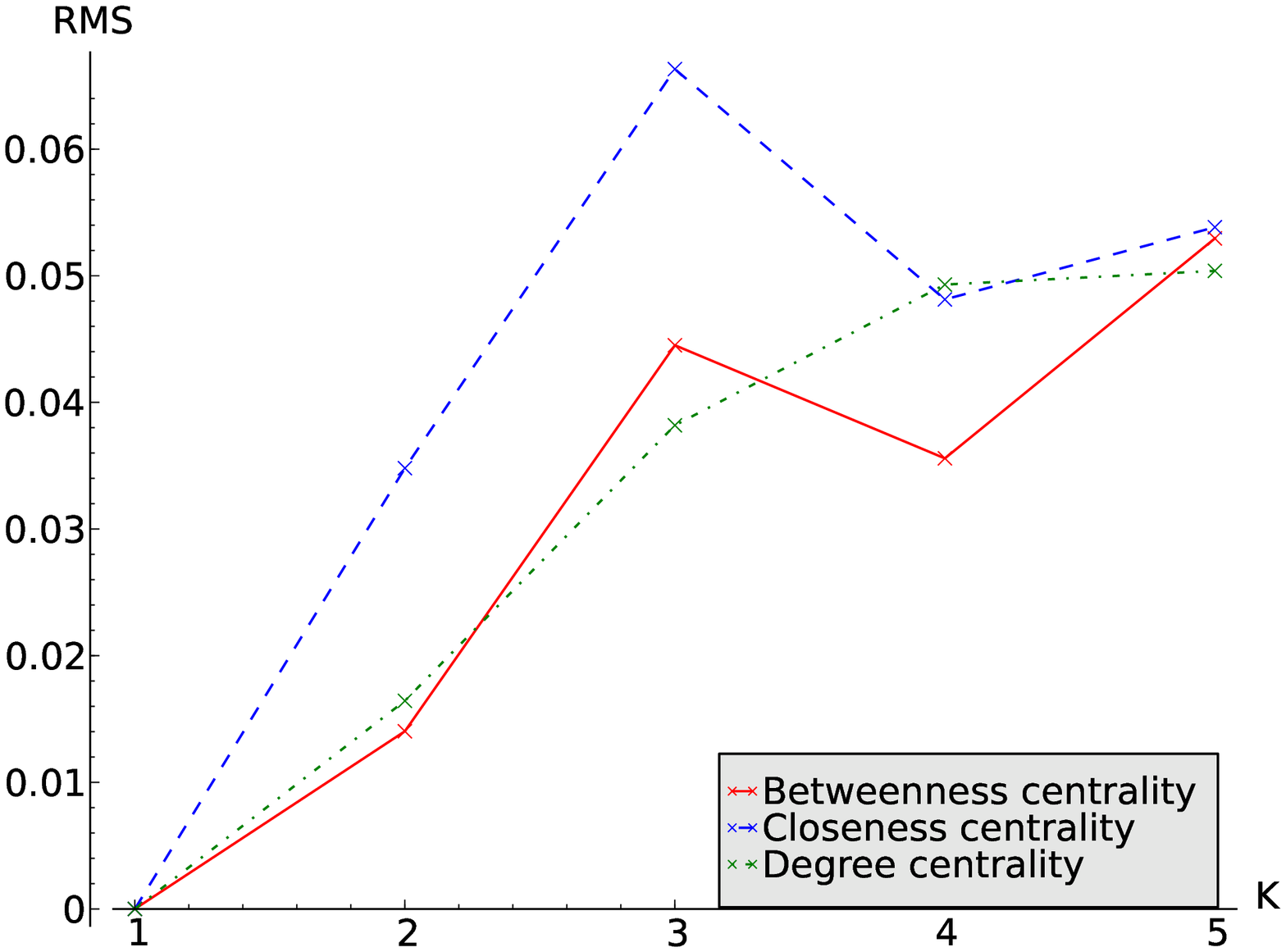}}~
	\subfloat[American College]{\label{fig:football-2}\includegraphics[width=\figureWidth]{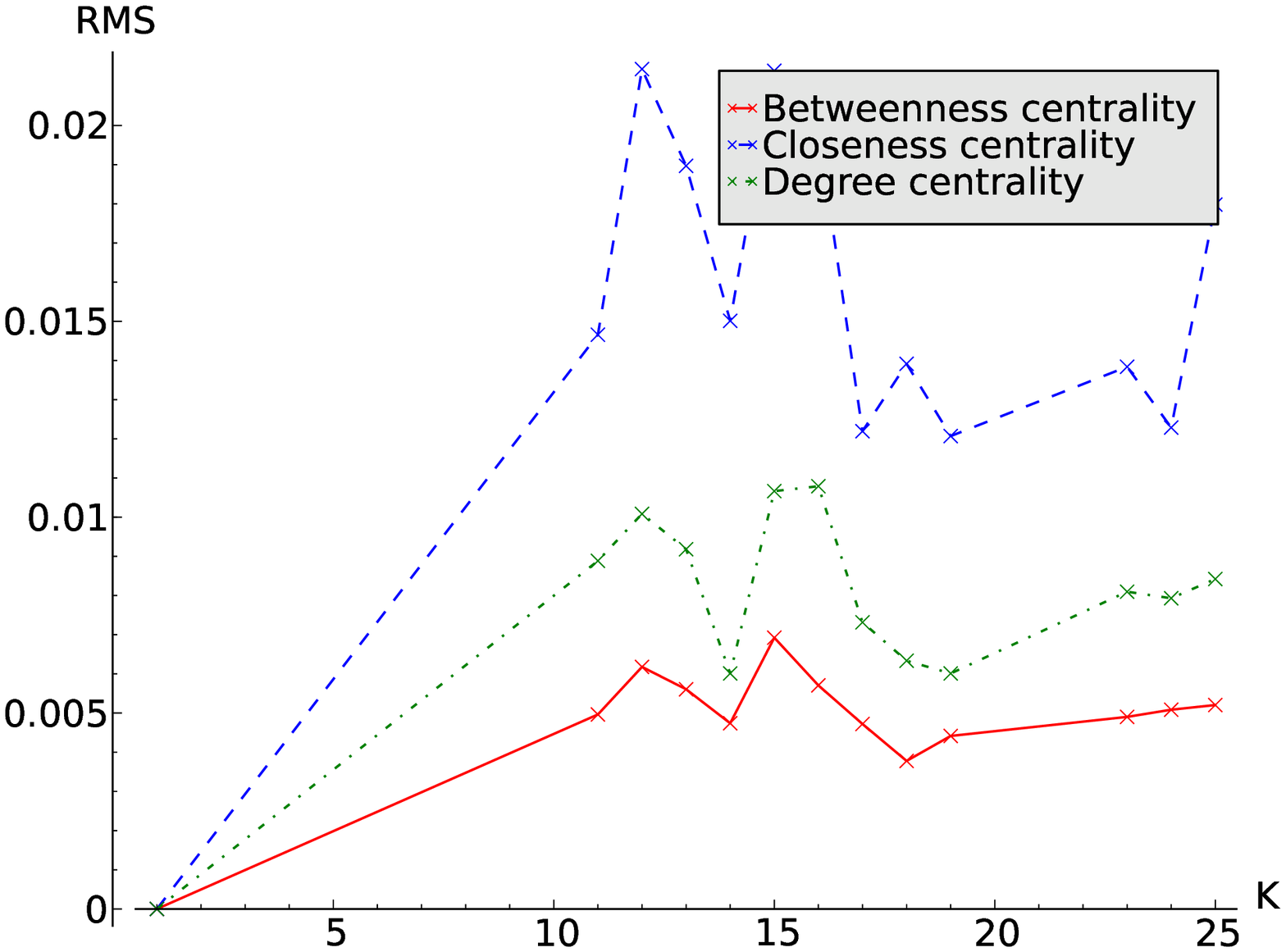}}~
	\subfloat[Jazz Musicians]{\label{fig:jazz-2}\includegraphics[width=\figureWidth]{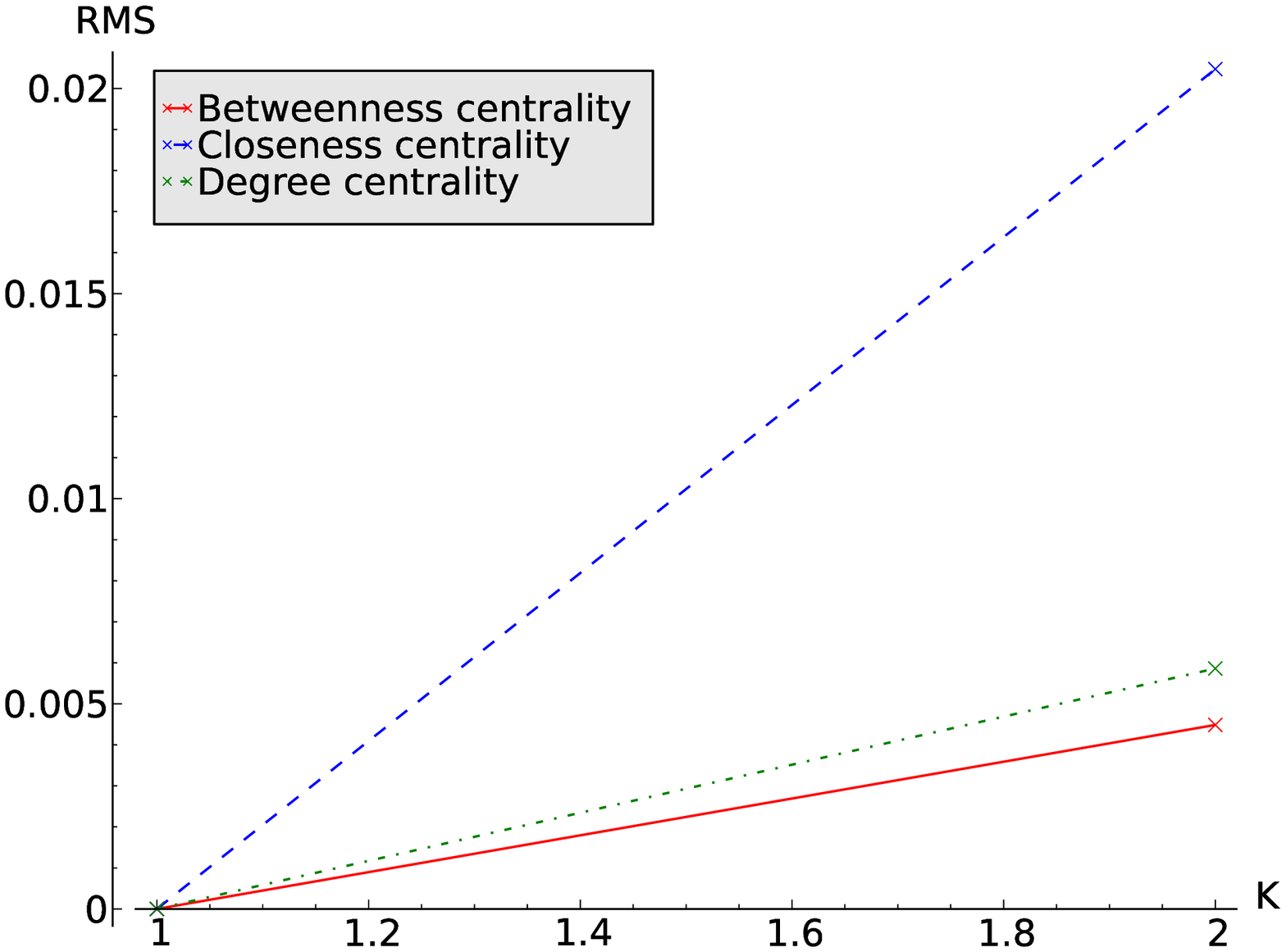}}~	
	\\	
	\subfloat[$Cand_{\mathcal{H}_{1}}$ (Zachary's Karate Club)]{\label{fig:karate-4}\includegraphics[width=\figureWidth]{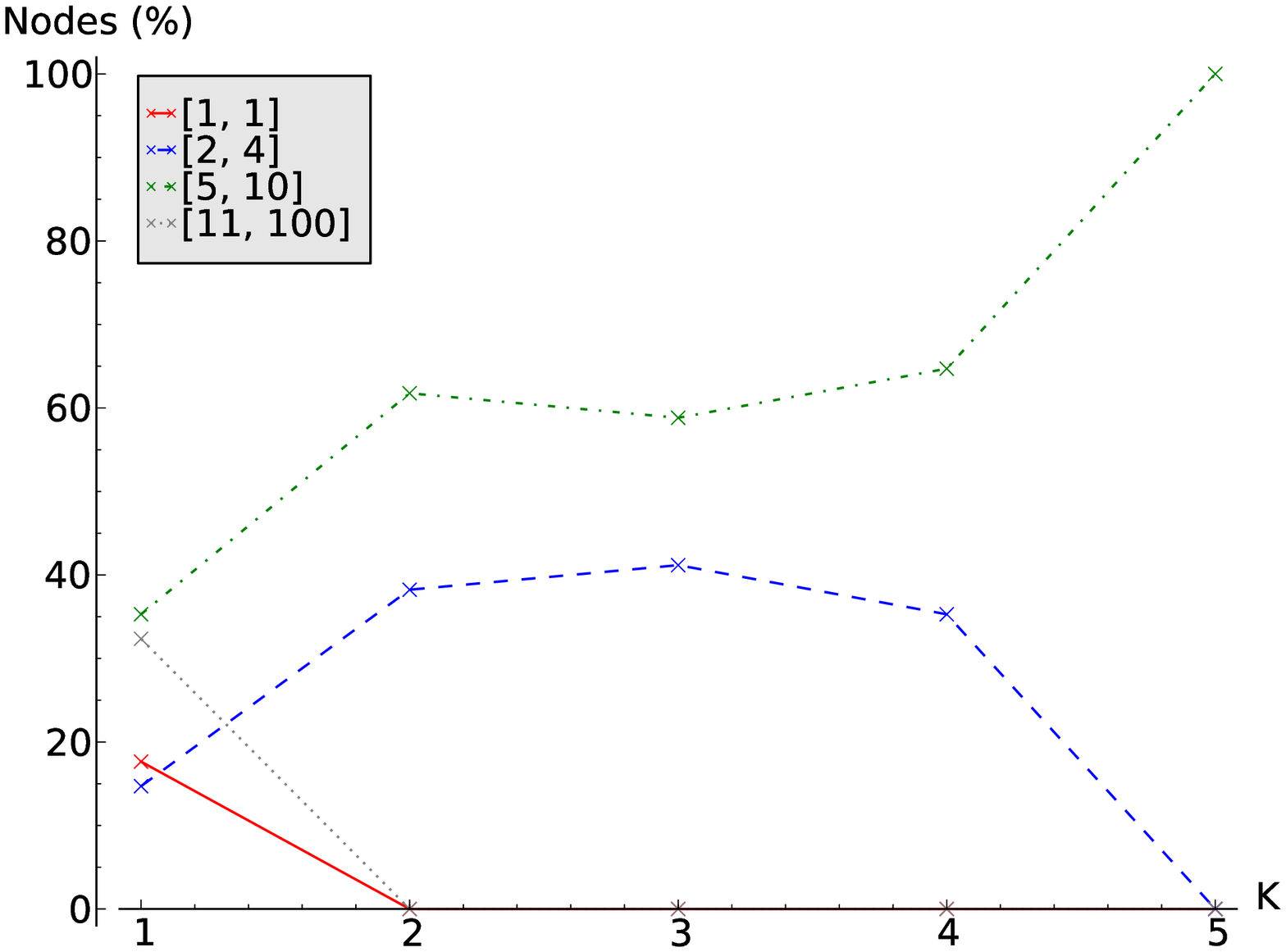}}~
	\subfloat[American College]{\label{fig:football-4}\includegraphics[width=\figureWidth]{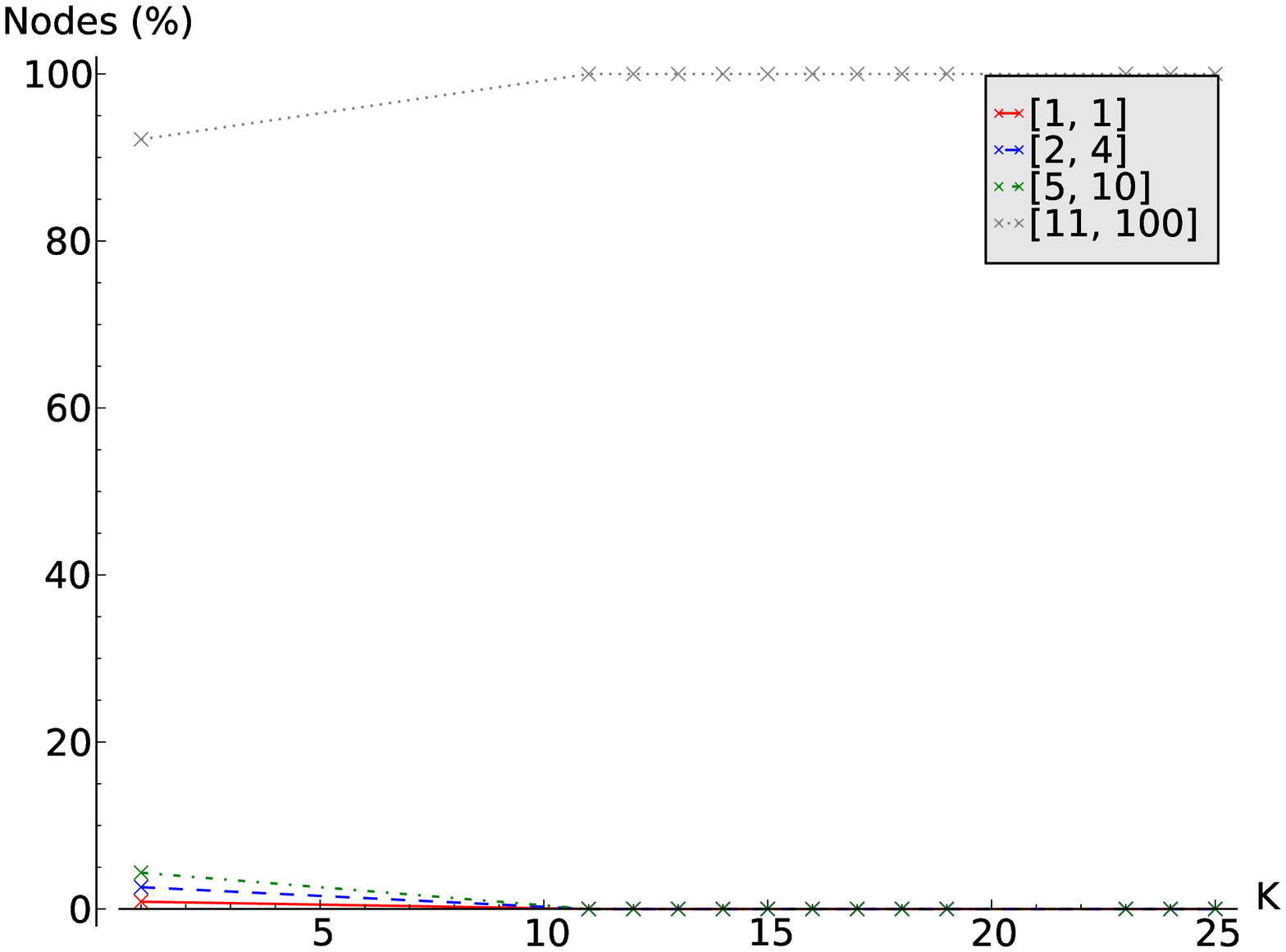}}~
	\subfloat[Jazz Musicians]{\label{fig:jazz-4}\includegraphics[width=\figureWidth]{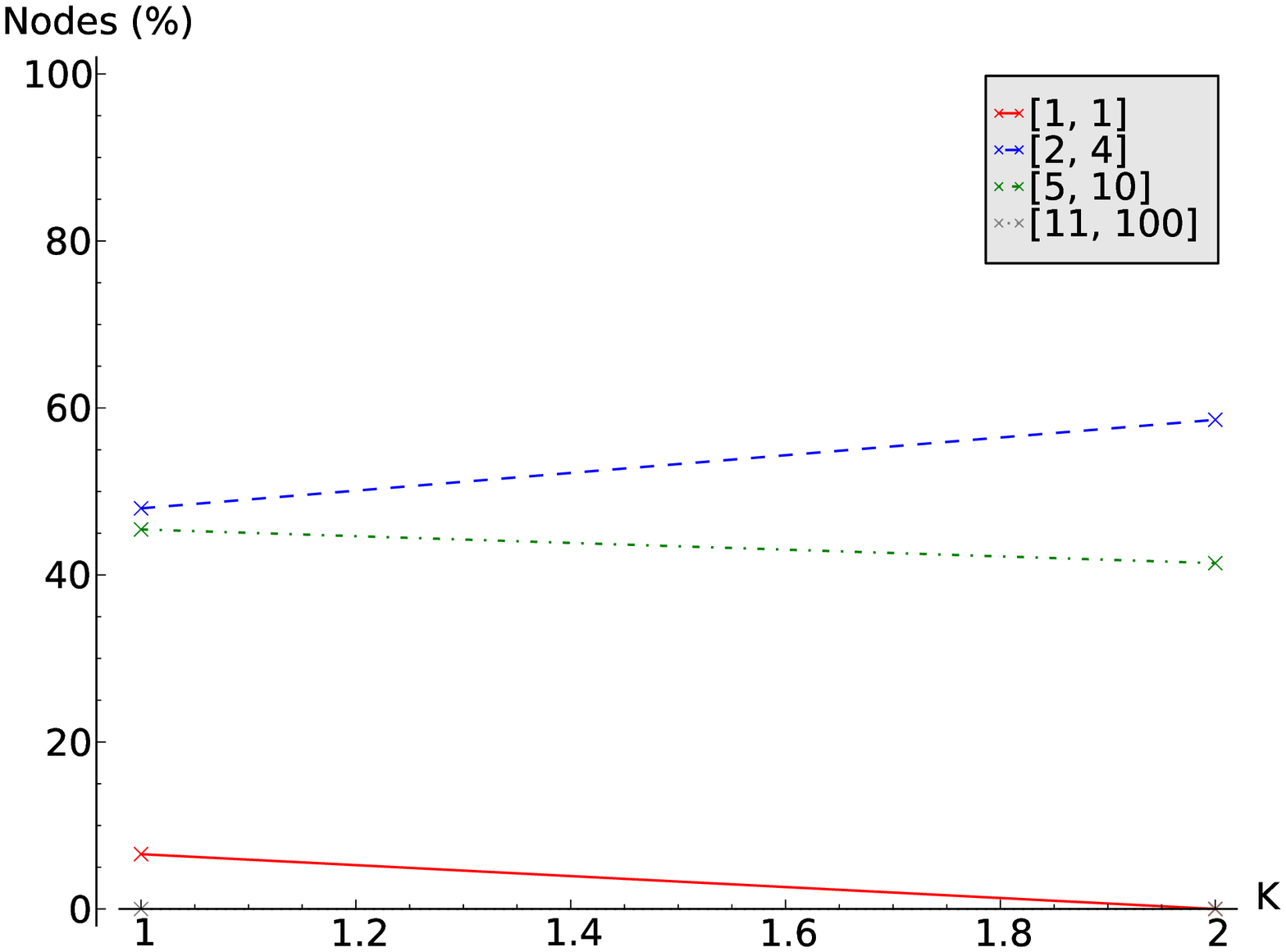}}~	
	\caption{Experimental results for selected graphs. The first row contains the degree histogram of original (grey) and anonymized (red) graphs. The second row contains the edge intersection as a function of $k$ (x axis). The following row presents the root mean square (RMS) of the three used centrality measures (i.e, betweenness, closeness and degree centrality) and finally, the fourth row shows the $k$-degree analysis by $Cand_{\mathcal{H}_{1}}$. The solid red line shows the percentage of nodes which can be directly re-identified (i.e, nodes with a unique degree value), the dashed blue line shows the percentage of nodes with high risk of re-identification (i.e, groups between 2 and 4 nodes with the same degree value), the dashed and dotted green line shows percentage of nodes with moderate risk of re-identification (i.e, groups between 5 and 10 nodes with the same degree value) and, finally, the dotted grey line shows percentage of nodes with low or very low risk of re-identification (i.e, groups of more than 11 nodes with the same degree value).}
	\label{fig:eaga-results-1}
\end{figure*}

The first dataset is a small social network with 34 nodes, 78 edges and a $k$-degree anonymity value equal to 1. EAGA algorithm anonymizes it to $k$ values equal to 2, 3, 4 and 5. Figure \ref{fig:karate-1} shows the original and $k=4$ anonymized degree histogram. As we can see, the degree histogram of original graph follows the power-law (total number of nodes exponentially decrease when degree value grows). Edge intersection is shown in Figure \ref{fig:karate-3}. An anonymized graph with $k=2$ the algorithm achieves an edge intersection value of 94.87\%, while this value descends to 79.49\% when $k$-anonymity value is equal to 5. Figure \ref{fig:karate-2} shows the RMS on the three centrality measures. Clearly, we want to keep these values closer to zero, since the less values, the less noise introduced on anonymized data. Though, the error increases as the $k$ value grows. Finally, we show the evolution of $Cand_{\mathcal{H}_{1}}$ as an extension to $k$-degree anonymity value, since this information allows us to see how nodes evolve in terms of re-identification during the anonymization process. As shown in Figure \ref{fig:karate-4}, the number of nodes which can be directly re-identified (solid red line) descends to zero when $k$ achieves a value of 2, while the number of nodes in high risk of re-identification (dashed blue line) does not fall to zero until $k=5$. When $k$ is equal to 5, all nodes are in moderate risk of re-identification (dashed and dotted green line).

The second dataset is a collaboration network with 115 nodes, 613 edges and a value of $k=1$. EAGA algorithm anonymizes it to a $k$ values in range 11-19, 23, 24 and 25. The degree histogram has experienced a few number of modifications in order to achieve a $k$-anonymity value of 19, as we can see in Figure \ref{fig:football-1}. Accordingly, this particular graph structure only needs a few edge modifications in order to achieve a great $k$-anonymity value. It is important to note that more than 93\% of edges remain the same on all anonymous graphs (96.25\% on graph with $k=19$ and 93.64\% on graph with $k=25$), as Figure \ref{fig:football-3} shows. Centrality measures, Figure \ref{fig:football-2}, show irregular perturbation on anonymous graphs. Finally, $Cand_{\mathcal{H}_{1}}$ is shown in Figure \ref{fig:football-4}, presents an important decrease on groups with direct, high and moderate re-identification risk until a zero value has been reached when the $k$-anonymity value is equal to 11. From this $k$ value on, all nodes are well-protected.

The third and the last dataset is a collaboration network with 198 nodes, 2,742 edges and a $k$-degree anonymity value equal to 1. This graph presents an average degree quite higher than others, close to 27 edges/node. The degree histogram in Figure \ref{fig:jazz-1} reveals the existence of two outlier nodes, with degree values of 96 and 100. It is important to underline that these two nodes are in danger of direct re-identification, but at the same time due to their important centrality they are key nodes for the graph structure (hubs). EAGA algorithm anonymizes the graph only to a $k=2$ value, due to the problem of the outlier values. It would be necessary to modify the two hubs of the graph in order to increase the $k$-anonymity value and, therefore, those hubs nodes would lose their centrality and produce high perturbation on anonymized data, reducing significantly the data utility. The anonymized graph with $k=2$ presents a high percentage of edge intersection (97.08\%), as we can see in Figure \ref{fig:jazz-3}, and a low error on centrality measures (Figure \ref{fig:jazz-2}), which indicates that small perturbations have been introduced on the anonymous graph. Finally, the $Cand_{\mathcal{H}_{1}}$ is presented in Figure \ref{fig:jazz-4}. It only points out a decrease on group of nodes with direct re-identification.

\section{Conclusions}
\label{sec:conclusions}

In this paper we have presented an algorithm for graph anonymization, based on edge modification approach in order to achieve a desired $k$-anonymity value on anonymized graph. The new algorithm, called Evolutionary Algorithm for Graph Anonymization (EAGA), is based on evolutionary algorithms to anonymize the degree sequence. The target of the evolutionary algorithm is to achieve a $k$-degree anonymous sequence and minimize the distance between the anonymous degree sequence and the original one, in order to preserve the data utility. The second step applies iterative edge swap to the original graph until the degree sequence is equal to the $k$-degree anonymous one. 

The results are favourable and indicate that the algorithm is able to provide anonymized graphs with a value of $k$-anonymity greater than the original one, and also keep a low perturbation level on anonymized data.

Many interesting directions for future research have been uncovered by this work. Firstly, recombination of parents as a generation process should be considered. Although this does not seem to improve the process, in-depth analysis would be of interest for future works. Secondly, other measures to quantify the perturbation introduced on anonymized graphs could be implemented and analysed. Thirdly, other types of graphs should be considered. For instance, directed or weighted graphs present new challenges to anonymization process.

\bibliographystyle{IEEEtran}

\end{document}